\documentclass[12pt,preprint]{aastex}
\usepackage{times}
\usepackage{amssymb}
\newcommand{\beq}{\begin{equation}}
\newcommand{\eeq}{\end{equation}}

\def\bar{\overline}
\def\OV{\overline{\bf V}}
\def\OB{\overline{\bf B}}
\def\bV{\overline V}
\def\emf{\overline{\mbox{${\cal E}$}} {}}
\def\emfb{\overline{\mbox{\boldmath ${\cal E}$}} {}}
\def\bbE{\bar {\bf E}}

\def\beq{\begin{equation}}
\def\eeq{\end{equation}}
\def\lsim{\mathrel{\rlap{\lower4pt\hbox{\hskip1pt$\sim$}}
    \raise1pt\hbox{$<$}}}
\def\gsim{\mathrel{\rlap{\lower4pt\hbox{\hskip1pt$\sim$}}
    \raise1pt\hbox{$>$}}}
\def\bfE{{\bf E}}
\def\bfJ{{\bf J}}

\def\bfA{{\bf A}}
\def\bfa{{\bf a}}
\def\bfe{{\bf e}}
\def\bfB{{\bf B}}
\def\bbJ{\bar {\bf J}}

\def\bB{\overline B}
\def\ts{\times}
\def\lb{\langle}
\def\rb{\rangle}
\def\curl{\nabla {\ts}}
\def\nt{\nabla\times}

\def\bbV{\bar {\bf V}}
\def\bfv{{\bf v}}

\def\bfj{{\bf j}}

\def\bfe{{\bf e}}

\def\bfb{{\bf b}}

\def\bfB{{\bf B}}

\def\bbB{\overline {\bf B}}
\def\bbA{\overline {\bf A}}
\def\nt{\nabla\times}
\def\div{\nabla\cdot}
\def\OB{\overline{\bf B}}

\begin{document}

\shorttitle{\textsc  Accretion  Disks and  Dynamos:  Toward a Unified Mean Field Theory}
\shortauthors{\textsc BLACKMAN }

\title{Accretion  Disks and  Dynamos:  Toward a Unified Mean Field Theory}
\author{Eric G. Blackman$^{1}$}
\affil{1. Dept. of Physics and Astronomy, University of Rochester,
  Rochester NY 14627, USA; blackman@pas.rochester.edu }

\begin{abstract}
Conversion of gravitational energy into radiation near stars and compact objects in accretion 
  disks and the origin of large scale magnetic fields in astrophysical rotators have often been distinct topics of  active research  in astrophysics.  In semi-analytic work on both problems it has been useful   to presume  large scale symmetries, which necessarily results in mean field theories;   magnetohydrodynamic turbulence  makes the underlying systems  locally asymmetric and highly nonlinear.     Synergy between  theory and simulations should  aim for the development of practical, semi-analytic mean field models  that capture the essential physics  and  can be used for observational modeling.    Mean field dynamo (MFD) theory  and alpha-viscosity accretion disc theory have exemplified such ongoing pursuits.  Twenty-first    century MFD theory has more nonlinear predictive power compared to 20th century MFD theory, whereas   alpha-viscosity accretion theory is still in a 20th century state.   In fact, insights from MFD theory are  applicable to accretion theory and  the two are really artificially separated pieces of what should ultimately be a single coupled theory.   I  discuss pieces of    progress  that provide clues toward a unified theory.  A key concept is that large scale magnetic fields can be sustained  via local or global magnetic helicity fluxes or via relaxation of small scale magnetic fluctuations, without appealing to the traditional kinetic helicity driver of 20th century textbooks.    These concepts may help  explain the formation of large scale fields that supply non-local angular momentum transport via coronae and jets in a unified theory of accretion and dynamos.  In diagnosing the role of helicities and helicity fluxes in disk simulations, it is important to study  each disk hemisphere separately  to avoid being potentially misled by the cancelation that occurs as a result of reflection asymmetry. The fraction of helical field energy in disks is expected to be small compared to the total field  in each hemisphere as a result of shear, but can still play a fundamental role in  large scale dynamo action.

\end{abstract}


\section{Introduction}


Large scale dynamo (LSD) theory  in astrophysics is aimed at quantitatively understanding the in situ physics of magnetic field  growth, saturation, and  sustenance  on time or spatial scales large compared to turbulent scales of the host.  The presence of magnetohydrodynamic  turbulence makes the subject particularly rich, highly  nonlinear, and exacerbates  the importance of numerical simulations.  However, large scale spatial symmetry and slow evolution of  large scale fields  compared to turbulent fluctuations  motivates  a mean field approach in which statistical, spatial, or temporal averages are taken  and evolution of the mean field studied (e.g. Moffatt 1978).

 An important goal of  semi-analytic mean field dynamo (MFD) theory  is to capture the essential  nonlinear physics simply enough for phenomenological models, but rich enough to account for the nonlinear saturation found in simulations  (for technical reviews see: e.g. Brandenburg \& Subramanian 2005; Blackman 2007).  
The 21st century has brought significant progress in this endeavor 
as the  growth and saturation of LSDs seen in   simple simulations seem to be reasonably matched by  
newer MFD theories  that follow the time dependent dynamical evolution of magnetic helicity (e.g. Blackman \& Field 2002). The modern 21st century MFDs represent  two or three scale theories
that capture the basic principles of the inverse cascade of magnetic helicity first 
derived in the  spectral  model of Pouquet et al. (1976), which revealed the fundamental utility 
of tracking magnetic helicity evolution to understand large scale field growth.

The existence of LSDs is not controversial: The sun proves that in situ LSDs  operate in nature since its large scale field  reverses   its sign over each half  solar cycle (e.g. Wang \& Sheely 2003;  Schrijver  \&  Zwaan 2000).   If the field were simply that frozen into to the ISM during initial formation, there would  be no reversals.  LSDs are also likely  operating in galaxies, despite whatever initial fields may have been seeded cosmologically because  supernova driven turbulent diffusion would otherwise diffuse the large field. 
Detailed modeling of the geometry of galactic large scale fields can be matched by mean field theories \cite{beck96,vallee04,shukurov02}, though incorporation of the nonlinear LSD principles of the past decade for idealized dynamos into more geometrically realistic dynamos of astrophysical rotators with realistic boundary conditions setting is an ongoing avenue of research (e.g. Sur et al. 2007,2009).  

Given the role of LSDs in stratified rotating  stars and galaxies,  
 their presence in turbulent accretion discs seems likely.
  Large scale fields likely play an integral role in the angular momentum 
 transport  for many real systems and this is evidenced by the presence of coronae and jets. 
 The global scale fields required by standard models of astrophysical jets (e.g. Blandford \& Znajek 1977; Blandford \& Payne 1982)  may be produced by an in situ LSD, although understanding the relative importance of in situ production vs. flux accretion  in that context has been a long-standing subject of research (e.g. Lovelace et al. 2009; Beckwith et al. 2011).
Non-thermal coronal luminosity from accretion engines also hints at the role of LSDs  because only large scale magnetic structures can  survive the buoyant rise to the corona
without being dissipated in the disk if the latter is turbulent (Blackman \& Pessah 2009).
Indeed, magnetohydrodynamic turbulence is  expected in accretion discs (e.g. Balbus \& Hawley 1998). If all of the large scale fields in the corona and jet were  produced in the disk, then
understanding the physics of coronal relaxation is an important part of assessing what fraction
becomes the global  jet fields and what fraction produces smaller loops that dissipate in the corona.
  
Amplification of large scale fields   is  now routinely seen in a wide range of accretion-disk-motivated simulations:  Localized stratified shearing box simulations indeed show evidence for large scale field reversing  LSD cycles  (Brandenburg et al. 1995; Lesur \& Ogilive 2008; Davis et al. 2010; Guan \& Gammie 2011).  Global simulations, now becoming more computationally feasible, 
    also show such LSD cycles and the  large
    scale magnetic fields either in coronae or jets
    (e.g. Beckwith et al. 2011, Sorathia et al. 2010, Tchekhovskoy et al. 2011, Romanova et al. 2011)
    though criteria for convergence (Hawley et al. 2011; Bodo et al. 2011) and sensitivity of
 the  dynamical importance of  associated vertical angular momentum transport 
    on the initial conditions (e.g. compare Penna et al. 2010 and Tchekhovskoy et al. 2011).
 must  be further investigated.  The stress gradients associated with the large scale fields
 in real systems may be significant contributors to   angular momentum transport that
 act in symbiosis with  with local turbulent transport.
  As will be  discussed later, an important point is that LSDs need not depend on the presence of kinetic helicity as a driver. In fact the electromotive force needed for LSDs     can be magnetic fluctuation driven (Blackman \& Field 2004) or even helicity flux driven (Blackman \& Field 2000a, Vishniac \& Cho 2001; Subramanian \& Brandenburg  2005)  These concepts are important for magneto-rotational instability (MRI) unstable systems (Balbus 2003) because magnetic fluctuations typically exceed the velocity fluctuations and differential shear can conspire to produce non-negligible   helicity fluxes (Vishniac 2009).
 
While LSD theory has always been presented with  explicit mean field theories that expose the approximations made, axisymmetric accretion disc models such as the Shakura Sunyaev  (Shakura \& Sunyaev 1973; hereafter SS73) viscous $\alpha_{ss}$ model are not always recognized as mean field theories by those using them to model spectra. 
 In this sense, 21st century LSD theory is more progressed than axisymmetric mean field accretion disc theory:
The $\alpha_{ss}$ disc prescription  does not yet offer predictive power for numerical simulations, whereas 21st century LSD theory  predicts the saturation of LSDs seen in a range of  simulations.  Ironically, because large scale fields likely play a fundamental role in accretion discs, both MFD and accretion  theory are really two faces of a single coupled theory. 

The subjects of LSD theory and accretion  involve the study of  intrinsically time dependent  non-equilibirum processes and  are in direct alignment  with the themes of the "Turbulent Mixing and Beyond" 2011 meeting:    Turbulent  "mixing" in the LSD context 
takes the form of  turbulent diffusion of large scale magnetic fields. Part of  the study of  LSDs  involves   understanding how   large scale  growth mechanisms compete with turbulent diffusion and what role  turbulent transport plays in establishing  large scale field geometries and cycle periods. 
In accretion disks, turbulent diffusion, or more generally, MHD mediated turbulent transport, has long been thought to be involved in at least some of the necessary extraction of angular momentum.  It is likely that the transport of large scale fields and transport of angular momentum are intimately linked in magnetized accretion disks. This collection of integrated topics continues to benefit from   symbioses between  large-scale numerical simulations, experiments, and basic theory.

Below I  first review  conceptual progress of 21st century LSD/MFD theory.
 I then discuss parallels between LSD theory and accretion disc theory.
Finally,  I discuss emerging connections  between the two 
how they should be unified into one.

 \section{21st Century LSD Theory}

For $\sim$ 50 years, 20th century textbook MFD theory (e.g. Moffatt 1978)
 lacked a  saturation theory to predict how strong the large scale fields get
before  quenching  via the back-reaction of the field on the driving flow. 
But substantial progress toward a nonlinear  mean field theory has emerged  in the 21st century  via a symbiosis between analytical and numerical work.  Coupling the dynamical evolution of magnetic helicity into the dynamo equations turns out to provide fundamental insight to the saturation seen in simulations. Whether this approach is the only way to understand the saturation is uncertain, but it has been fruitful.

The connection between magnetic helicity and large scale dynamos is first evident in the classic EDQNM spectral model of helical MHD turbulence in Pouquet et al. (1976). 
In Kleeorin \& Ruzmaikin (1982)   an equation that  couples the time evolution of small scale magnetic helicity to the LSD was given but not solved.  A time-dependent two scale LSD model
that incorporates the principles of Pouquet et al. (1976) was derived and solved 
in  Blackman \& Field (2002), and  Field \& Blackman (2002) showed that solutions matched simulations of Brandenburg (2001).
For a variety of different discussions  see  Brandenburg \& Subramanian (2005);  Blackman (2007); K\"apyl\"a et al. 2008) and below.

\subsection{Globally Helical }

Most  work in LSD/MFD theory has focused on systems that are initially globally reflection asymmetric (GRA).   In such systems,  pseudoscalars, such as the hemispherically averaged product of angular velocity and density gradient imposed by the initial setup
facilitate e.g. a kinetic helicity  pseudoscalar, common to  standard textbooks (Moffatt 1978; Parker 1979)
``$\alpha_{dyn}$ effect'' of mean field dynamos.   However,  21st century MFD theory has revealed that  magnetic helicity and its evolution provides a more fundamental  unifying quantity for all MFDs. 
 The initial GRA conditions and the potential pseudoscalars that arise are important
 primarily because they provide one way to sustain  
an electromotive force aligned with the mean magnetic field. This alignment is a source
of magnetic helicity or magnetic helicity flux.

 There are two  classes of GRA LSDs.
The first is flow dominated helical dynamos (FDHD) which apply
 inside of astrophysical rotators. Here the initial energy is dominated by velocity flows and thermal energy.  The magnetic field may grow larger than the fluctuating velocities on some scales, but
 the total kinetic energy in all forms remains larger than the total magnetic energy in FDHDs.  

The FDHDs are potentially linked to coronae via boundary terms 
where  a second type of LSD, the magnetically driven helical dynamo (MDHD) can operate.  A MDHD can also be  described as dynamical magnetic relaxation (Blackman \& Field 2004) 
toward the "Taylor" state (Taylor 1986).
Note that the Taylor state is the lowest energy state to which a magnetically dominated configuration evolves subject to boundary conditions.
For a closed system,  the state is characterized by  the magnetic helicity having migrated to the largest scale available.  In an accretion disc, the  MDHD would  characterize coronal relaxation of magnetic structures (fed by helical fields from below)
into  larger (even jet mediating) scales in a magnetically dominated environment.
In this respect, the MDHD in disc-corona or star-corona  interfaces are  analogous to laboratory plasma dynamos (\cite{bellan,jiprager})
 that occur in reverse field pinches (RFPs) and Spheromaks. The corona becomes the "internal volume" of these devices and the disc or stellar surface acts as  the injecting boundary.

LSDs always involve some helical growth of the large scale field, which is coupled
to a helical growth of smaller  scale fields of opposite sign
and/or a compensating helicity flux when boundary conditions allow.
When small scale magnetic or current helicity evolution is coupled to the large scale 
field growth, the simplest 21st century dynamo reveals that the mean field `dynamo $\alpha_{dyn}$ becomes the difference between kinetic helicity and 
current helicity:  For an $\alpha^2_{dyn}$ type FDHD simulated in a closed box (Brandenburg 2001), 
the current helicity builds up as the large scale field 
grows and quenches the FDHD, in accordance with MFD predictions (Blackman \& Field 2002;). 

The effect of boundary or flux terms can vary depending on the sign and 
relative flux of small and large scale magnetic helicities. For a closed system   the buildup of small scale magnetic (and current) helicity 
is necessarily accompanied by small scale magnetic helicity buildup with the opposite sign,
which quenches the LSD to at most a rate determined by the dreadfully slow resistive dissipation of small-scale helicity.
If this "catastrophic" quenching occurs before enough large scale field is grown in the fast growth regime, 
 a preferential flux of  helicity of the opposite sign through a realistic astrophysical boundary
is desirable to  alleviate this quenching  and sustain further fast fast growth.  For GRA systems,
in each hemisphere, the large scale helicity builds with one sign and the small scale with the opposite sign so an outward flux of  small scale helicity is desirable to sustain the electromotive force (Blackman \& Field 2000ab; Vishniac \& Cho 2001; Sur et al. 2007; K\"apyl\"a et al 2008).
It is possible that even if the field grows to acceptable magnitudes without boundary fluxes,
the cycle period becomes resistively limited.   
In this case the  helicity fluxes can unclog the cycle.
I will come back to the role of helicity fluxes in section 3.2.

For an MDHD, the system is first dominated instead by the 
current helicity and a growing kinetic helicity can act as the back-reaction 
(Blackman \& Field 2004; Park \& Blackman 2012). 
Both simple FDHDs and MDHDs are accessible within the same formalism, all unified by tracking magnetic helicity evolution, and aided by thinking of the field as ribbons rather than lines
(Blackman \& Brandenburg 2003).

Interestingly, although most observations probe coronare, most work on astrophysical dynamo theory has focused on the FDHD and not the MDHD.

\subsection{Not Globally Helical}


LSD action has also been observed in non GRA simulations 
(e.g. Yousef et al 08; Lesur \& Ogilvie 08; Brandenburg et al. 2008). 
The minimum global ingredients for this class of LSD seem to be 
 shear  plus turbulence  (Yousef et al. 2008).
The non GRA LSDs grow large scale fields on scales larger than the outer scale of turbulence but smaller than the global scale. Although there is no GRA,  in regions where the large scale field is coherent regions, there is a field aligned electromotive force (EMF), and thus  an intermediate scale source of magnetic helicity that may switch signs between coherence regions and globally averages to zero.  The most promising explanation for the Yousef et al. (2008) simulations
  seems to be  that of Heinemann et al. (2011) which involves fluctuations in the electromotive
  force. This is not unrelated to the implications of a stochastic $\alpha$ effect 
(e.g. Vishniac \& Brandenburg 1997 [stochastic $\alpha_{dyn}$ effect]).
When there are no helical fluctuations in the EMF and no pseudoscalars,
shear alone does produce an LSD  (Kolekar et al. 2012).
I would also point out that in sub-global patches where any mean field grows and the EMF is finite
for a given time span, its alignment with the local mean magnetic field means it is a source of large scale magnetic helicity locally.   In general,  the absence of GRA does not imply the irrelevance of magnetic helicity and/or its flux, in local subregions of a volume for which the global flux may vanish.





\section{Why Tracking  Magnetic Helicity in  Large Scale Dynamos Provides Insight}

The time evolution of the 
 mean field  $\OB$ is given by the  induction 
equation (Moffatt 1978; \citet{kr80}) 
\beq
{\partial\OB\over \partial t} = -c\curl\bbE,
\label{2.4a} 
\eeq
where
\beq
c\bbE=-\OV\times \OB  - \lb\bfv\ts\bfb\rb+
{\nu_M }\nt\OB,
\label{2.4aaa} 
\eeq
where the magnetic diffusivity $\nu_M\equiv {\eta c^2\over 4\pi}$  and 
$\eta$ is the resistivity.
The  turbulent electromotive force $\emfb\equiv \lb \bfv\times \bfb \rb $  can  be expanded in terms of powers of spatial and temporal derivates of $\OB$ and $\OV$ (or mean shear).  The lowest order contributions to $\emf$ from the mean magnetic field that lead
to exponential growth and diffusion (and are the only contributions in the absence of a mean
velocity) are given by
\beq
\emf_i = \alpha_{ij}{\bB}_j-\beta_{ijk}\partial_j{\bB}_k.
\label{2.5aa}
\eeq

Using Maxwell's equations, the definitions  ${\curl \bfA }\equiv \bfB$, and $\bfJ\equiv {c\over 4\pi} \curl \bfB$, along with vector identities, the equation for the evolution of magnetic helicity density $H^M\equiv \lb \bfA\cdot\bfB \rb $ is  (e.g. Bellan 2000)
\beq
\partial_t( \bfA\cdot\bfB) 
=-2\nu_M(\bfJ\cdot\bfB)
-\div( 2\Phi \bfB + \bfA \ts \partial_t\bfA),
\label{4a}
\eeq
where $\Phi$ satisfies $\bfE = \nabla \Phi - {1\over c}{d\bfA\over dt}$.
Following the analogous procedure for the mean and fluctuating
quantities, and using (\ref{2.4aaa}) and ((\ref{2.5aa}) gives for the large and small
scale quantities $H_1^M\equiv \lb\bbA\cdot\bbB\rb$ and $H_2^M\equiv \lb \bfa\cdot \bfb \rb$ respectively
\beq
\partial_t(\bbA\cdot\bbB)
=2\emfb\cdot\bbB
-2\nu_M \bbJ\cdot\bbB
-\div( 2{\overline \Phi}\ \bbB + \bbA \ts \partial_t\bbA)
\label{5a}
\eeq
and 
\beq
\partial_t\overline{ \bfa\cdot\bfb}
=-2\emfb\cdot\bbB-
2\nu_M\overline{\bfj\cdot\bfb}
-\div(\overline{2{\phi} \bfb} + \overline{\bfa\ts \partial_t\bfa}).
\label{6a}
\eeq

Standard MFDs necessarily involve generation of  mean magnetic helicity. To see this, note that the minimalist versions of such dynamos invoke
$\alpha_{ij}= \alpha_{dyn}\delta_{ij}$ and $\beta_{ijk}=\nu_{M,T}\epsilon_{ijk}$ where $\alpha_{dyn}$ 
is a  pseudoscalar  and $\nu_{M,T}$ acts as a mean field scalar diffusion coefficient. 
The growth of the large scale field then involves
a finite $\emfb\cdot\OB$, which provides an equal and opposite
source term to the large scale and small scale magnetic helicity as seen from
(\ref{5a}) and (\ref{6a}). 
 In solving (\ref{2.4a}), standard 20th century textbook theory
focuses on an equation for the large scale field from which one can write
an equation like (\ref{5a}). But 20th century dynamo theory does not couple in
the evolution of the small scale magnetic helicity contained
in (\ref{6a}).

In the simplest 21st century LSDs which  consolidates aspects of the spectral study of Pouquet et al. (1976) into two  scales  (e.g. Field \& Blackman 2002; Blackman \& Field 2002; Blackman \& Brandenburg 2002),  $\alpha_{dyn}\sim -\tau (\lb \bfv\cdot\curl \bfv \rb -\lb\bfb\cdot\curl\bfb\rb)\equiv  \alpha_{kin} + \alpha_{mag}$, where $\tau$ is of order a correlation time of the dominant fluctuations.
In  20th century versions of helical dynamos (Moffatt et al. 1978) only $\alpha_{kin}$ contributes to $\alpha_{dyn}$.  The contribution $\alpha_{mag}$ is proportional to the small scale  current helicity, which in turn is proportional to $H_2^M$ in the Coulomb gauge for a closed system. Therefore the time evolution of $\alpha_{dyn}$ is directly coupled to the time evolution of magnetic helicity. If  $H_1^M$  grows initially from $\alpha_{kin}$, then  $H_2^M$ grows of opposite sign which in turn quenches $\alpha_{dyn}$. 

\subsection{ Kinetic vs. Magnetic Fluctuation Driven LSDs}

Comparisons between theory and numerical experiments  have validated the basic theoretical framework of 21st century LSD just described. The simplest experiments are those of the so called $\alpha^2$ dynamo in which isotropically driven kinetic helicity is injected into a periodic box at intermediate wave number $k\sim 5$, and the $k=1$ large scale field grows (Brandenburg 2001). 
Agreement between the basic theory and simulations of such kinetically forced (KF) 
dynamos has been shown in (Field \& Blackman 2002; 
see also Brandenburg \& Subramanian 2005; Park \& Blackman 2012a).  
The  insight gained from  comparison of the basic
$\alpha^2$ LSD theory to the simulations  is that when kinetic helicity is forced and drives large scale  helical magnetic fields, the small scale helical field grows because magnetic helicity is largely conserved up to resistive terms. The buildup of the small scale current helicity associated with the small scale field then off-sets the  driving kinetic helicity and quenches the LSD.  Ultimately
the slightly faster resistive decay of the small scale field than the large scale field 
leads to an asymptotic steady state with net helicity associated with that of the large scale sign.

However, this physical mechanism and the equations that describe it also motivate consideration of the complementary case of driving the initial system with current helicity rather than kinetic helicity.  This was investigated analytically in Blackman \& Field (2004) where it was found that driving the system so as to supply initial small scale  current helicity  produced a magnetically forced (MF) analogue to the $\alpha^2$ dynamo.   
 Because the MF case is   driven by magnetic fluctuations, there is no "kinematic" regime in the sense that the magnetic fluctuations exceed kinetic fluctuations  from the outset. Thus the analogue of the kinematic  regime for the MF $\alpha^2$ case would be  a "magnematic" regime where the small scale velocity fluctuations are small. 

  Numerical experiments in which   the system is forced in the induction equation with an electric field that drives small scale magnetic helicity  (Brandenburg et al. 2002; Alexakis et al. 2006:  Park \& Blackman (2012b) show that indeed the large scale field grows in response.
 Park \& Blackman (2012b)  compared the MF theory to MF numerical experiments
   and show that  the buildup of the kinetic helicity  explains the end of  magnematic regime and quenches the MF analogue of the $\alpha^2$ dynamo.  Thus kinetic and current helicity reverse their roles between the KF and  MF cases.
   The growth of the large scale field in the MF vs. KF cases has another important distinction:
    Because of magnetic helicity conservation, the large scale field grows with the same sign as that injected for the MF case. In the KF case, the injection of kinetic helicity does not change the net magnetic helicity and so the large and small scale magnetic helicities grow with opposite signs.
  
 The ratio of magnetic to thermal energy $\beta_p \ge 1$  at all times in the simulations of  Park \& Blackman (2012b) and so the simulations thus demonstrate the existence
   of a  $\beta_p>1$ magnetic fluctuation driven dynamo. 
Although only the simplest MF LSDs have been considered by employing  no shear or rotation and periodic boundaries, the basic concept of a LSD that is driven by small scale magnetic flucations rather than kinetic fluctuatons in $\beta_p >1$ environments may have important conceptual relevance to stellar and accretion disk dynamos. In those cases, it should be noted however  that the Coriolis force, anisotropic turbulence, and buoyancy conspire to allow still other contributions to the EMF.  (e.g. Brandenburg \& Subramanian 2005).

Large scale dynamos are now commonly seen in accretion-disk-motivated simulations
(Brandenburg et a.. 2005, Davis et al. 2010, Gressel 2010..41G, K\"apyl\"a and Korpi, 2011; Guan \& Gammie 2011).  The MRI operating in these simulations produces turbulence that drives an LSD
and the magnetic fluctations typically exceed the kinetic fluctuations.
Agreement between LSD theory and simulation indeed seems to require
consideration that  the driver of the LSD growth is not the kinetic helicity term but may be the current helicity term (Gressel 2010).  This would correspond to a $\beta_p>1$ magnetic fluctuation (or current helicity) driven dynamo. Note that the total flow kinetic energy also still dominates the magnetic energy because the Keplerian shear contains the dominant energy in the system. Moreover, 
in any thin disk, the helical fraction of the large scale field is expected to  be subdominant in magnitude but still essential for diagnosing the LSD mechanism.


Note that in the $\beta_p<<1$ limit, MF LSDs have long been studied in the
laboratory context (e.g. Bhattacharjee \& Hameiri 1986, Ji \& Prager 2002). The direct analogy to these low $\beta$ MF LSDs may occur in astrophysical coronae. We come back to this point later.

\subsection{Magnetic Flux Driven Dynamos}

An emerging frontier in  21st century astrophysical dynamo theory is the recognition of
the importance of open boundaries  because the electromotive force can be sustained
exclusively by boundary terms. These boundary terms can continue to drive the EMF
independent of the aforementioned quenching  that arises because of competition between kinetic and current helicities. 

To introduce the basic point
 consider the steady state limit of 
Consider now the limit of (\ref{5a}) and 
(\ref{6a}) in which the time evolution and resistive terms
are ignored, but the divergence
terms are kept. We then have respectively, 
\beq
0=2\lb\emfb\cdot\bbB\rb
-\div\lb{\overline\Phi} \bbB + \bbE\ts \bbA\rb
\label{26}
\eeq
and 
\beq
0=
-2\lb\emfb\cdot\bbB\rb
-\div\lb{\phi} \bfb + \bfe\ts \bfa\rb
\label{27}
\eeq
Combining these two  equations reveals that  the fluxes
of large and small scale helicity through the system
boundary are equal and opposite  in a steady state. In principle, the flux of small scale helicity
can sustain an $\bbB$-aligned EMF,  $\emfb_{||}$, that grows the large scale field of oppositely signed helicity provided
that there is a source of  energy for the amplified magnetic field.

For a fixed 
amount of magnetic helicity, the energy in a magnetic structure  is  minimized when this helicity
evolves to as large a scale as possible.  Thus if the rate of helicity fluxes of small and large scales are equal, more  small
scale energy and current helicity would be leaving per unit time than that associated with the large scale field.  If, for example, there is a source of EMF within the rotator from kinetic helicity, then a  flux of small scale magnetic helicity can help alleviate the backreaction from the current helicity and allow the dynamo or cycle period to avoid quenching (e.g. Blackman \& Brandenburg 2003; K\"apyl\"a et al. 2010).


In general, both the time derivatives and the  flux terms 
in equation (\ref{5a}) and (\ref{6a}) should be included dynamically.
In the context of the Galaxy, Shukurov et al. (2006)   
solved the mean field induction equation for $\bbB$ 
assuming that $\emfb_{||}$
satisfies 
$\partial_t \emfb_{||}=0$,
As discussed above,  $\emfb_{||}$, involves the difference between the kinetic and current 
helicities which can be related to
small scale magnetic helicity in the Coulomb gauge.
(Shukurov et al. (2006) formally use a gauge invariant
helicity density, derived by Subramanian \& Brandenburg (2006)
to replace the use of the magnetic helicity density but
the key role of the boundary terms is conceptually independent of this.)  , 
Shukurov et al. (2006)  solve the induction  equation for $\bbB$, 
(which depends on $\emfb_{||}$ and thus $\lb\bfa\cdot\bfb\rb$)
and Eq. (\ref{6a}) for $\lb\bfa\cdot\bfb\rb$.
They replace the divergence term in (\ref{6a})  with one of the form 
$\propto \nabla\cdot (\lb\bfa\cdot\bfb \rb \bbV)$, where $\bbV = (0,0,\bV_z) $ is the mean
velocity advecting the small scale helicity out of the volume.
This mean velocity also appears in the induction equation for $\bbB$, 
highlighting that the  loss terms in the small scale helicity equation also
imply advective loss of mean field.
This approach exemplifies the general concept
(Blackman \& Field 2000ab; Blackman 2003) 
that a flow of small scale helicity toward the boundary may help
to alleviate the backreaction on  the small  scale  magnetic helicity on the 
kinetic helicity, when the latter drives the dynamo in  $\emfb_{||}$.  
If however,  if $\bV_z$ is too large, it may
carry away too much of the desired  mean field which the dynamo is trying to grow
in the first place.   

A different time dependent dynamo that includes boundary terms, maintains the time dependence in (\ref{5a}), but implicitly assumes that equation (\ref{6a}) reaches a steady-state has also been
studied (Vishniac \& Cho 2001; (See also Brandenburg \& Subramanian 2005 for further review).  This explicitly   incorporates the role of shear into the helicity flux and has been argued to be important in explaining angular momentum transport in accretion disks (Vishniac 2009,)
Indeed, the  role of some kind of shear dependent helicity flux may alleviate what appears to be 
the asymptotic vanishing of MRI turbulent stresses in unstratified shearing boxes at large magnetic Reynolds number (Vishniac 2009; K\"apyl\"a and Korpi 2011) and has been shown to alleviate
catastrophic quenching in convection driven dynamos as well 
 (K\"apyl\"a et al. 2010).   


A subtle aspect of the helicity density equations (\ref{5a}) and (\ref{6a}), and 
thus for computing dynamos with open boundaries and helicity fluxes, 
is the  issue of gauge invariance. In the absence of boundary flux terms, all terms are manifestly
gauge invariant.  In the present of flux terms, the gauge invariance is broken (e.g. Berger \& Field 1986).  Subramanian \& Brandenburg (2006) carefully construct a generalized local helicity density that reduces to the above equations in the absence of flux terms 
and has the same form in the presence of flux terms but with a redefinition of $H_1^M$,  $H_2^M$ and the flux terms that avoids the use of the vector potential. The lowest order terms in their
definition are similar to what is obtained from the Coulomb gauge but the corrections make
each individual  terms  in the equation gauge invariant with a local physical meaning.
That important construction puts the physical discussion of helicity density evolution  equations here and elsewhere on  firm ground, particularly if one seeks to measure a local  flux density as a physical quantity.  However, if one uses the helicity evolution equations as intermediary equations in a particular gauge and  can convert back to the magnetic field, the resulting magnetic field will always be gauge independent.  

Numerical simulations may be easier to carry out in particular gauges and
the specific choices of boundaries may be troublesome for specific gauges.
In this context, as emphasized by Hubbard \& Brandenburg (2011), 
 the specific Vishniac-Cho (VC) flux is  typically computed in  the Coulomb gauge.
Hubbard \& Brandenburg (2011) have argued that the Coulomb gauge is problematic
numerically for periodic shear flows. Appealing to the fact that flux divergences can be gauge invarian
even when fluxes are not, and working in a  numerically appealing shear-periodic gauge  
they find that for the unstratified periodic box conditions previously though to incur a VIshniac-Cho flux, the vertical flux vanishes even though the divergence of the total
flux does not. They argue that the VC flux is a horizontal flux and not a vertical one,
and therefore would not be useful to large scale dynamos as a small scale helicity release valve.
It may be that the enhanced turbulent diffusion of small scale helicity (e.g. Blackman 2003; Hubbard \& Brandenburg 2010)
 plays a more prominent role. 
 
 A further subtlety is that  Hubbard and Brandenburg (2012) have pointed out  that shear-periodic
 boundaries lead to  helicity fluxes that that flip sign at what should otherwise be shear periodic
 boundaries.  (This is a bit troubling in the analogous way that vertically stratified periodic boxes
 have to set gravity to zero at the top and bottom boundaries to be sensible).
  In a horizonally averaged analysis, they purport to avoid dealing with these fluxes by working with a set of equations that avoids  solving the time evolution of the small scale magnetic helicity.  In this approach,  the flux terms hidden,  working "behind the scenes"  to prevent the dynamo from quenching prematurely.   However, with  
 K. Subramanian (K. Subramanian 2012, personal comm.) we have found that there seem to be
flux terms that  are dropped form their alternative set of equations both in their horizontally averaged case and for more general extensions which do not seem to be easily argued to vanish so 
more work is needed to assess whether helicity fluxes work can really be thought to work "behind the scenes" or need to be explicitly calculate.  The issue is of practical importance, while the alpha-squared dynamo can grow sizeable mean field and saturate at that value even without helicity fluxes,
the question of whether an $\alpha-\Omega$ dynamo requires helicity fluxes to reach a steady state
with a sizeable mean field before that field rapidly decays is directly related to this issue.
(e.g. compare Sur et al. 2007; to Hubbard and Brandenburg 2012).
There is more to be done to quantify helicity flux driven LSDs and
corroborate theory and numerical simulation.

Magnetic helicity fluxes into  astrophysical coronae, and the subsequent evolution
of such structures   are directly  analogous to the  evolution
 magnetically dominated laboratory plasma  subject to the injection of magnetic helicity
 as in   a Spheromaks (e.g. \cite{bellan}, Hsu \& Bellan 2002) or reversed field pinches (Ji \& Prager 2002). Such relaxation process have been studied as helicity flux driven dynamos in magnetically dominated plasmas  (e.g. Bhattacharjee \& Hameiri 1986).

The  experiment of Hsu \& Bellan 2002 
provides an analogy to helical loops of flux rising
into an astrophysical corona from its rotator below.
The loops coalesce at the symmetry axis and form
a magnetic tower. For large enough helicity injection, 
the tower can break off a Spheromak blob from the kink instability. 
An astrophysical corona can  be modeled as an aggregate 
of magnetic loops subject to reconfiguration, reconnection, and relaxation 
 (Blackman \& Field 04; Uzdensky \& Goodman 2008).
The helicity flux to the corona in astrophysical rotators
acts  as a seed for this subsequent  magnetic relaxation (or MDHD)
Field lines that open up global may be those which enable jets.

\section{Accretion  Theory  and  the Connection to LSDs}

\subsection{Standard accretion theory is  a mean field theory}

Gaseous accretion discs around astrophysical rotators have long been studied as
a source of luminosity  (see Treves et al. 1989; Frank et al. 2002).
In astronomy,  a  practical 
semi-analytic formalism to be incorporated into emission models for comparison with observations is a desired goal.   Practical accretion models
in standard texts  (e.g. Frank et al. 2002)   
 based on SS73  employ the 
Navier-Stokes equation and replaces the microphysical viscosity with a turbulent viscosity
without explaining the theoretical complexity and assumptions underpinning 
this  bold procedure.    The replacement of the microphysical viscosity by a turbulent viscosity
is itself a turbulent closure and warrants a formal derivation analogous to that 
of the MFD theory coefficients.  Closures approximate an otherwise infinite set of nonlinear turbulence equations by an approximation that facilitates a finite set.

That the SS73  model is both axisymmetric AND involves a turbulent viscosity implies that the theory is a mean field theory because local axisymmetry never applies
on the scale of turbulent fluctuations.  That accretion disc theory is a mean
field theory is addressed in a subset of theoretically oriented literature 
 (e.g. Balbus et al. 1994; Blackman 1997; Balbus \& Hawley 1998; 
 Ogilvie 2003 Pessah et al. 2006; Hubbard \& Blackman 2009). 
MFD theory has traditionally focused on the magnetic induction equation as the "primary equation" of the theory and accretion disc models focus on the mass conservation and angular momentum transport equations as the "primary equations" of the theory with the coupling to the magnetic field
often ignored or "swept" into "$\alpha_{ss}$".
At present, a formal identification of the minimal set of 
transport coefficients that best predicts the angular momentum transport
 seen in  simulations and can be used for practical spectral modeling is lacking.

 MFD theory is more progressed than mean field accretion theory because 
 there is a  semi-analytic MFD theory that agrees with  a set of nonlinear simulations, as discussed earlier.  Recall from section 2 that   MFD theory has presently 2 tensor  coefficients in common use (the $\alpha_{ij}$ and $\beta_{ijk}$ coefficients which are  used even in semi-analytic dynamo models to match observed field structures in stars or in galaxies. 
 Dynamical equations  that incorporate magnetic helicity evolution and employ "minimal $\tau$" closure (e.g. Blackman \& Field 2002, see also Snellman et al. 2009 applied to shear flows) capture  MFD saturation.  In accretion theory however, there are few models that use anything other than  the single  diffusion coefficient of the SS73 formalism.  The minimum properties that 
turbulence must have to transport angular momentum outward are more subtle than what purely isotropic turbulence provides, and it is likely that more than one transport coefficient  is needed (Pessah et al. 2006; Hubbard \& Blackman 2009). Progress toward capturing the saturated stress of  the magneto-rotational instability (MRI) as a mean field model using a closure similar to the "minimal $\tau$"  approximation and going beyond
the $\alpha_{ss}$ disc model may be slowly emerging (Ogilvie 2003; Pessah et al. 2006).

Although the  magneto-rotational instability (MRI) has emerged 
as a leading candidate for local angular momentum transport in  accretion discs (e.g. Balbus \& Hawley 1998; Balbus 2003),  we do not yet have quantitative
scalings for the measured transport coefficients that can be used in practical disk emission models.
Simulations have not converged to the extent  that robust scalings of transport coefficients can be  extracted (Hawley et al. 2011).  Perhaps even more substantially, since most MRI "accretion disk" simulations have neither disks (they are cartesian boxes) nor accretion (there is no torque) even if we were able to obtain converged stresses, we still would be deficient in determining torques.
The latter really requires more global simulations.
In particular, there is a need to determine the relative fraction of vertical torque that transports angular momentum vertically compared to radially, and the subsequent fraction of that vertical transport that divides into corona vs. jet.

Most first generation MRI simulations (except Brandenburg et al.1995)
also did not use explicit viscosity or magnetic diffusivity.
Indeed more recent work does show that for unstratified simulations
there is a  magnetic Prandtl number dependence
(Fromang et al. 2007; Lesur and Longaretti 2007).
The dimensionless transport parameter   $\alpha_{ss}$ depends strongly on the box size and the strength of the initially imposed weak mean field strength for unstratified 
boxes with imposed mean fields (Pessah et al. 2007), though
though this dependence is reduced for stratified boxes (Davis et al. 2010; Shi et al. 2009).  
Interestingly, $\alpha_{ss}$ varies $\sim 4$ orders between simulations, 
but $\alpha_{ss}\beta_p$, where $\beta_p$ is the ratio of thermal to magnetic pressure, 
is nearly a constant (Blackman et al. 2008).  This is maintained in global simulations as well
(e.g. Romanova et al. 2011) 

The mean field approach  of  Pessah et al. (2006)  hints at what minimal set of coefficients might be required in a more robust but still practical model of MRI mediated accretion discs.  
The approach better matches  the stress dependence on rotation profile than the $\alpha_{ss}$ of SS73   in numerical simulations of unstratified boxes with an initial mean field
 (Pessah et al. 2008).
.


\subsection{Role of  Large Scale Fields: Clues  from  MRI Simulations}

Due to computational limitations,  much of  the early computational study of the promising MRI has focused on the local manifestation of the instability  as a source  of local turbulence  that can transport angular momentum radially outward  (e.g. Balbus \& Hawley 1998, Balbus 2003)
Understanding the local physics is important but  even classic  paradigms  of accretion discs  
involve large scale  magnetic loops or global fields
  (Lynden-Bell 1969; FIeld \& Rogers 1993; Tout \& Pringle 1992; Johansen and Levin 2008)
  or outflows  (Blandford \& Payne 1982; 
K\"onigl 1989,  Tager \& Pellat 1999; Colgate et al. 2001; Caunt \& Tagger 2001
K\"onigl  et al. 2010) as primary transport agents.
There have also been  a few efforts to incorporate the dynamics of vertical transport and its effect on the radiation spectra in semi-analytic models of accretion disks
  (Kuncic \& Bicknell 2004,2007; Dobbie et al. 2009).

Mesoscale shearing box simulations (Davis et al. 2010; Guan \& Gammie 2011; SImon et al. 2012). 
  and global accretion disk simulations (Penna et al. 2010, Beckwith et al. 2011, Sorathia et al. 2010, Tchekhovskoy et al. 2011, Romanova et al. 2011) are becoming more numerous but
 convergence is still an issue (Hawley et al. 2011; Bodo et al. 2011) and  compromises still have to be made.
 For example,   simulations may  include stratification but  not explicit resistivity and viscosity (or vice versa).
  The choice of periodic vs. open vertical boundaries simulations is another distinction with practical
  implications.   For global  disk simulations, the computational time limitations typically require choosing an initial seed magnetic field that is relatively large.  Assessment of which results from  simulations  depend on initial conditions and boundary conditions needs to be further investigated.
The torques rather than the stresses are the most relevant quantities to determine for angular momentum transport.  

Stratified local or meso-scale simulations do show less dependence on resolution and box size  than unstratified solutions (Pessah et al. 2007; Bodo et al. 2008; Davis et al. 2010)
though are seemingly still not absolutely converged (Guan \& Gammie 2011; Hawley et al. 2011)
 Since real systems are stratified, the weaker dependence of the saturated state on box properties is  step forward, but with stratification comes magnetic buoyancy and coronae. Large scale structures have the longest turbulent diffusion times and are most likely to survive the buoyant rise (Blackman \& Pessah 2009). Thus if stratification is important
then understanding what radial and vertical scales are large enough to fully capture
the dominant magnetic stress remains a question for global simulations. 

Global mean  toroidal magnetic fields that last many orbit times and exhibit sign reversals
are now seen routinely in local shearing box simulations 
(Brandenburg et al. 1995, Lesur \& Ogilvie 2008, Davis et al. 2010; Guan et al. 2011; Oishi and MacLow 2011; Simon et al. 2012) and
in global simulations
(Penna et al. 2010 ,Beckwith et al. 2011, Sorathia et al. 2010, Tchekhovskoy et al. 2011, Romanova et al. 2011).   The cycles are evident in  in stratified simulations with and without periodic vertical boundaries (Oishi and MacLow 2011)
, and  even in unstratified simulations with periodic vertical boundaries (Lesur \& Ogilvie 2008).  The patterns indicate a LSD operating contemporaneously with the small scale dynamo
that produces magnetic fluctuations.   Understanding the extent to which the large scale fields transport angular momentum vertically and non-locally is a desired pursuit. In all of these cases however, there is at minimum a Coriolis force which presents a pseudo-scalar ${\bf \Omega}\cdot {\hat {\bf z}}$ and anisotropy which can source an EMF, and in turn , sources magnetic helicity growth.
(See also R\"udiger and Pipin 2000;  Rekowski et al. 2000; Brandenburg and Subramanian 2005).

It is important  to note several principles in analyzing shearing box simulations for LSD action and the magnetic helicity.  FIrst, when there is a mid-plane, any pseudoscalars or reflection asymmetric quantities  reverse sign. When any helicities are computed by averaging over the entire box
 one expects that such quantities will vanish (regardless of the boundary conditions). In addition, the helicities may have opposite signs on different scales even within a given hemisphere. Thus one should  separately compute helicity  spectra  in each hemisphere to analyze their role.    This is  in addition to evaluating any boundary terms.  In this context note that Oishi \& MacLow (2011) found that the stratified periodic and stratified open boundary simulations are similar with respect to LSD action which is an important result. However, since their analysis of the possible role of magnetic helicity did not involve the separate  analysis of helicity in each hemisphere we cannot yet rule out the role magnetic helicity dynamics as being fundamental to the LSD action in the simulations they compare.  Even with a closed box, local helicity fluxes may  be important.  A further point is  the helical fraction of the magnetic field is not a  measure of the importance of magnetic helicity in that we expect shearing boxes to be have predominantly non-helical fields due to shear. The needed LSD process by which some  toroidal field is converted to poloidal field and sustains the field against decay may still depend fundamentally on a dynamics  that tracking magnetic helicity evolution may help explain.

.

\section{Combining MFD and Accretion Theory}

Here I outline the basic ingredients of a "strawman" paradigm for sufficiently ionized, non-self-gravitiating  accretion engines that combine the MRI and LSDs. There are two flavors of scenarios: (1) those led by an EMF supplied by pseudocalars fluctuations and aided by helicity fluxes or (2) those fully dominated by helicity fluxes with pseducalar contributions to the EMF small.  Fluxes are the common ingredient to both paradigms.

For a disk that becomes unstable to  MRI driven turbulence, stratification and rotation in a real system then imply a source of reflection asymmetry.  Because the MRI produces magnetic fluctuations that dominate velocity fluctuations one pseudoscalar of  relevance is the current helicity. Provided there is sufficient current helicity from these magnetic fluctuations,  a current helicity driven LSD can result.   Such an LSD that is driven by  current helicity fluctuations (rather than kinetic helicity) is conceptually consistent with a generalization the $\beta_p>1$ magnetic fluctuation driven dynamo discussed in section 3.1 to include shear  and may be a plausible model  for LSDs observed in some MRI simulations  (Gressel 2010). 

The combination of the small scale magnetic fluctuations driven by the MRI and the  current helicity driven LSD  will produce  a spectrum of magnetic energy and a spectrum of magnetic stress.  
Magnetic structures large enough to buoyantly rise 1-2 density scale heights before being shredded
by turbulent diffusion  are the structures that form the corona (Blackman \& Pessah 2009).  
The  vertical angular momentum transport likely results only from the structures   that satisfy the buoyancy condition so there  would be a 1 to 1 correspondence between the fraction of total magnetic stress associated with these structures and the  ratio of vertical to total stress.   
(Derivation of the correlation length of magnetic fields from MRI simulations is a useful goal in this context such as in  Simon et al. 2012).
The local radial stress would be  due to  magnetic structures who scale  is below the critical scale for buoyancy.  
It is  desirable to extract the ratio of vertical to total stress from simulations and compare to the corresponding properties of the magnetic  stress spectra for structures above and below
the   critical scale for buoyancy. It would also be of interest to determine the non-local radial stress from large scale loops anchored at different radii. The torques associated with the vertical and non-local stresses could be comparable to the local radial torque even if the stress ratio is small since vertical gradients can be larger than radial gradients.

Magnetic structures  large enough scale to survive the buoyant rise to  the  corona also have magnetic helicity and even likely have some amount of both positive and negative sign by analogy to the sun (Blackman \& Brandenburg 2003; Brandenburg et al. 2011 ).
The flow of magnetic helicity to the corona from the disk is capture by a  flux term  that can sustain $\emfb_{||}$ as discussed in section 3.2. When  fields reach the corona, they can further relax via  dynamical  magnetic relaxation (MDHD) in what would be a  $\beta_p<1$ environment.
There will be some competition between relaxation to large scales and dissipation
when loops of  opposite sign reconnect, and ultimately some coronal spectrum
will result (see also Uzdensky \& Goodman 2008, who did not however invoke magnetic helicity conservation).  The largest scale fields will mediate the jets and the coronal loops will be agents of
 hard X-ray emission (at least in AGN disks) upon reconnection (e.g. Field \& Rogers 1993).
Whether the helicity fluxes that are accompanied by a flux of magnetic energy help sustain the pseudocalar driven EMF as in e.g. Sur et al. 2007 or by themselves sustain the LSD as in e.g. 
Vishnical (2009) is unclear at present.

As just described, the transport of angular momentum involves
both radial and vertical components. The local radial component would leads to 
thermal disk emission and the vertical and (and some non-local radial) would contribute to jets and coronae. Both  large scale fields and the small
scale fields are then important . A generalized mean field accretion disk theory must therefore start with equations for the mean and fluctuating components of the magnetic field and  the velocity and employ a 
suitable closure  to reduce the set of mean field equations to closed form, leaving  a finite set of transport coefficients.  In this way, corona, jet, and disk emission can all be coupled into a single theory.  I think that finding such a theory is tractable.  At first one may have to compromise aspects of a  dynamical theory for parameterization informed by simulations, but the ultimate goal would be to determine the underlying principles behind each transport coefficient.
One can envision both 1-D (vertically integrated, azimuthally averaged) and
2-D (only azimuthally averaged) theories. 
Note  that  issues such as the competition between diffusion of global  scale
magnetic fields in accretion discs, advection of flux, and amplification of flux are all part of the same question: how do mean magnetic fields evolve in accretion discs when the appropriate minimal set of  mean field transport coefficients are obtained for both the coupled  LSD and accretion dynamics?
 Ultimately, mean field accretion disc theory and LSD theory are  coupled.

Some aspects of the coupled evolution of mean velocity field in accretion flows and
mean magnetic fields in dynamo theory have been studied in semi-analytic models. However, principles of 21st century LSD theory have yet to be included in this coupling.
Campbell \& Caunt (1999) combined 20th century MFD theory with the SS73
accretion theory  closure to include local and large scale stresses and 
Campbell (2003) studied the vertical and radial accretion disk structure coupled to an LSD supplied magnetic field and magnetized wind.
Kuncic \& Bicknell (2004,2007) have derived a mean field MHD accretion theory that includes radial and vertical transport and have calculated the effect on the thermal disk spectrum and can accommodate the limit in which all of the transport is vertical (pure Poynting flux  accretion solution) . They have not included LSD theory or a detailed coronal + jet spectal model. Quataert \& Narayan (1999) included spectra from both disks and winds for advection dominate flows but  without including the MHD dynamics or an LSD theory. 
The SS73 type viscosity (via MRI) has also been used  as a closure for the turbulence from which  semi analytic MFDT is then derived to produce large scale fields in discs  to power outflows (e.g. Tan \& Blackman 2004). 20th century MFD theory has also been applied to MRI stratified simulations to explain observed cycle periods (Brandenburg et al. 1995;  Brandenburg \& Donner 1997; also Lesur \& Ogilvie 2008; Davis et al. 2010; Guan \& Gammie 2011). 

There is much more semi-anlaytic work to be done. In fact, as simulations become more and more sophisticated, the opportunities for analytic and semi-analytic work increases.


\section{Conclusions}

For any turbulent MHD  system in which there is both turbulence as well as
ordered large scale patterns,  there  exists an averaging and appropriate closure 
that dynamically  captures essential physics and key principles in a useful framework.
It may not be easy to find this closure  but the question should not be  "is mean field theory is correct?" but    "do we have the correct mean field theory?"
Progress in large scale dynamo theory is farther along that accretion theory in this regard; 
mean field dynamo theories now exist that make correct predictions of the nonlinear saturation of
simple large scale dynamos but we do not yet have  a theory for accretion disks that has either identified the complete set of transport coefficients, or predicts their time evolution.    
Noteworthy progress this  goal is the model described for periodic box MRi simulations in Pessah et al. (2008).  However, the ultimate desired theory needs to couple mean field accretion theory with mean field dynamo theory in such a way that the vertical and  large scale angular momentum transport,  evidenced by coronae and jets, are captured along with small scale local transport within the disk.

It is plausible that the MRI, operating in a stratified rotator with open boundaries, generates
a magnetic spectrum which has both:
(1) small scale MHD turbulence with magnetic structures below the scale for which buoyancy
beats turbulent shredding. These structures transport angular momentum radially
and account  for dissipation within the disk; (2) larger scale magnetic structures for which buoyancy beats turbulent diffusion and that  transport angular momentum to the coronae.
Present simulations suggest that large scale helical dynamos operate  within
accretion disks,  driven in part by current helicity fluctuations produced by  the MRI. 
The  flux of magnetic helicity from disk to corona can help sustain the large scale dynamo within
disk and feed the disk corona with magnetic fields.   A  fraction of the magnetic fields
that make it to the corona relax to open up and mediate  jets.  I think that it will be possible to develop a mean field theory with a finite set of useful transport coefficients
that can accommodate all of these ingredients and in fact still be practical for use by observational
modelers.

\acknowledgements
Thanks to K. Subramanian, O Gressel,   F. Naumann, and R. Penna, J. Oishi, and K. Park  for discussions.
I acknowledge NSF grants  AST-0406799, AST-0406823, and NASA grant ATP04 -0000-0016 and the LLE at UR. Thanks to the organizers of the "Turbulent Mixing and Beyond" meeting in 2011 in Trieste, Italy  for a stimulating meeting.


\end{document}